# Controlling colloidal phase transitions with critical Casimir forces

Van Duc Nguyen[1], Suzanne Faber[1], Zhibing. Hu[2], Gerard H. Wegdam[1] and Peter Schall[1]

[1]*Van der Waals-Zeeman Institute, University of Amsterdam, Amsterdam, The Netherlands*

[2]*Department of Physics, University of North Texas, Denton, Texas, U.S.A.*

**The critical Casimir effect provides a thermodynamic analogue of the well-known quantum mechanical Casimir effect. It acts between two surfaces immersed in a critical binary liquid mixture, and results from the confinement of concentration fluctuations of the solvent. Unlike the quantum mechanical effect, the magnitude and range of this attraction can be adjusted with temperature via the solvent correlation length, thus offering new opportunities for the assembly of nano and micron-scale structures. Here, we demonstrate the active assembly control of equilibrium phases using critical Casimir forces. We guide colloidal particles into analogues of molecular liquid and solid phases via exquisite control over their interactions. By measuring the critical Casimir particle pair potential directly from density fluctuations in the colloidal gas, we obtain insight into liquefaction at small scales: We apply the Van der Waals model of molecular liquefaction and show that the colloidal gas-liquid condensation is accurately described by the Van der Waals theory, even on the scale of a few particles. These results open up new possibilities in the active assembly control of micro and nanostructures.**

The critical Casimir effect provides an interesting analogue of the quantum mechanical Casimir effect [1-4]. Close to the critical point of a binary liquid, concentration fluctuations become long-range, and the confinement of these long-range fluctuations gives rise to critical Casimir interactions. This offers excellent new opportunities to achieve active control over the assembly of colloidal particles [5-10]. At close distance, solvent fluctuations confined between the particle surfaces lead to an effective attraction that adjusts with temperature on a molecular time scale. Because the solvent correlation length depends on temperature, temperature provides a unique control parameter to control the range and strength of this interaction [3]. The advantage of this effect is its universality: as other critical phenomena, the scaling functions depend only on the symmetries of the system and are independent of material properties, allowing similar interaction control for a wide range of colloidal particles [9]. Such interaction control offers fascinating new opportunities for the design of structures at the micrometer and nanometer scale with reversible control.

Here, we demonstrate the direct control of equilibrium phase transitions with critical Casimir forces. Through exquisite control of the particle pair potential with temperature, we assemble colloidal particles into analogues of molecular liquid and solid phases, and we visualize these phases in three dimensions and on the single particle level. Because of the similarity of the critical Casimir and molecular potentials, these observations allow insight into molecular liquefaction at small scales. In 1873, Van der Waals presented a universal model for the condensation of gases: Gas-liquid condensation reflects the competition between the energy cost for compression of the particles against their entropic pressure, and the energy gain from the condensation of the attractive particles. Van der Waals obtained a universal equation for the condensation of gases by accounting for the average molecular attraction and the finite molecule volume.

While for molecular gases, experimental confirmation of this relation remained indirect, our colloidal system allows direct microscopic observation and measurement of the liquid condensation. We measure important parameters of the condensation process: the particle pair potential, the equilibrium densities of liquid and gas phases, and relative amount of gas and liquid. These microscopic observations allow us to test the Van der Waals model at small scales. We find that the model describes the particle condensation in our colloidal system remarkably well, even on the scale of a few particles, indicating that this relatively simple mean-field model may also apply to describe the condensation of nanoparticles in the formation of nanostructures.

**Results**

We induce critical Casimir interactions by approaching the phase separation temperature, $T_c$, of the binary solvent; we tune these interactions continuously in a temperature range of $\Delta T = 0.5$ to $0.2$°C below $T_c$. By using a new optically and density-matched colloidal system (see Materials and Methods), we can directly follow individual particles deep in the bulk of the system. The convenient temperature control allows us to induce remarkable transitions in the colloidal system: Sufficiently far below $T_c$, the particles are uniformly suspended and form a dilute gas phase (Fig. 1a). However, at $\Delta T = 0.3$°C below $T_c$, the particles condense and form spherical aggregates that coexist with a low-density colloidal gas (Fig. 1b). We observe that within the aggregates, particles exhibit the diffusive motion characteristic of liquids (see supplementary video S1). This is confirmed by tracking the motion of the particles, and comparing their mean-square displacement with that of the particles in the dilute gas in Fig. 1c. The mean-square displacement grows linearly in time, similar to that of the molecular motion in liquids. Moreover, it grows slower than that of the gas particles, indicating the confinement of the particles in the dense liquid environment. We conclude that the particles form a condensed phase analogous to that of molecular liquids. When we raise the temperature further to $\Delta T = 0.2\,°C$, the particles inside the aggregates form an ordered face-centered cubic lattice (Fig. 1d): the colloidal liquid has frozen into a crystal. These observations provide direct analogues of gas-liquid and liquid-solid transitions, driven by critical Casimir interactions. Because of the exquisite temperature dependence of these interactions, the phase transitions are reversible: the crystals melt, and the liquid drops evaporate when the temperature is lowered below the characteristic thresholds (see supplementary video S2). Such reversible control offers new opportunities to guide the growth of perfect structures from colloidal building blocks [12]. The exquisite temperature dependence and reversibility allows precise control over the growth and annealing of perfect equilibrium structures, in analogy to the growth of atomic materials.

To elucidate the gas-liquid condensation in more detail, we show three-dimensional reconstructions in Fig. 1e and f, where we distinguish gas and liquid particles by their number of nearest neighbors [11]; red spheres indicate particles with five and more neighbors, and small blue spheres indicate particles with less than five neighbors. Interestingly, at $\Delta T=0.5°C$, small clusters of red particles appear and disappear, indicating spontaneous fluctuations in the density of the colloidal gas (Fig.1e). These density fluctuations become more pronounced as the temperature approaches $T_c$ (see supplementary video S1). At $\Delta T=0.3°C$, after 10 min, red spheres form stable liquid nuclei (Fig. 1f) that grow to become large drops. After 30 min, we measure the density of particles in the liquid and gas phases from the number of red and blue particles per volume, and obtain $\rho_{liq}=3.2\mu m^{-3}$ and $\rho_{gas}=0.05\mu m^{-3}$, respectively.

We elucidate the mechanism that drives the condensation of particles by measuring the particle pair potential directly from the density fluctuations in the gas. The potential of mean force, $U_{mf}$, is related directly to the pair correlation function, $g(r)$ that indicates the probability of finding two particles separated by $r$, via $g(r) \sim exp[-U_{mf}(r)/k_BT]$ [13]. For dilute suspensions, $U_{mf} \approx U$, and the pair potential is obtained directly from the measurement of $g(r)$. We acquire 3000 images of particle configurations of the colloidal gas to determine the average pair correlation function at different temperatures, which we show in Fig. 2a, inset. An increasing peak arises at $r \sim 2.7r_0$ as the temperature approaches $T_c$, indicating increasing particle attraction. The resulting pair potentials $U(r)/k_BT = -\ln g(r)$, shown in Fig. 2a, demonstrate an increasing attractive minimum as the temperature approaches $T_c$. This minimum reflects the competition between the critical Casimir attraction, and the screened electrostatic repulsion of the particles. We consider the total pair potential $U = U_{rep} + U_{attr}$, where $U_{rep}(l) \approx A_{rep} \exp(-l/l_{rep})$ is the screened electrostatic potential associated with the charged particles [14] and $U_{attr}$ is the attractive critical Casimir potential, which we approximate by $U_{attr}(l) \approx -A_{attr} \exp(-l/l_{attr})$ [3]. Here, $A_{attr} \sim 2\pi r_0 k_B T/l_{attr}$ is the amplitude and $l_{attr}$ the range of the critical Casimir attraction, and $l = r - 2r_0$ is the distance between particle surfaces. We determine the electrostatic repulsion from the measurement at room temperature, where critical Casimir interactions are negligible. Excellent agreement with the data is obtained for $A_{rep}=4.3\,k_BT$ and $l_{rep}=0.36\,r_0$ (black circles and solid line in Fig. 2a). The attractive critical Casimir potential is then determined from the best fit of $U$ to the measured potentials; the resulting critical Casimir potential is shown in Fig. 2b, and the values of $A_{attr}$ and $l_{attr}$ are shown in the inset. The data indicates that $l_{attr}$ grows linearly with temperature, while the amplitude $A_{attr} \sim 1/l_{attr}$, as expected for the critical Casimir force between two spheres [7]. The range of the attractive potential at $\Delta T=0.3K$ is $l_{attr}=80$nm, of the order of, but a factor of 2 larger than the solvent correlation length $\xi=40$nm determined from independent dynamic light scattering measurements. We associate this difference with the fluffiness of the NIPA particles that can give rise to an increased range of the attraction.

**Discussion**

The precision of these measurements allows us to make direct comparison with continuum models. The widely used Van der Waals equation of state provides a universal model for molecular gas-liquid condensation. To describe attractive molecules of finite size, Van der Waals modified the ideal gas equation of state $PV=Nk_BT$ that relates the pressure $P$, the volume $V$, the number of molecules $N$ and the temperature $T$, to $(V-Nb)\left(P+\dfrac{N^2 a}{V^2}\right)=Nk_BT$ [15]. The finite molecule size reduces the volume accessible to the molecules by $Nb$, where $b$ indicates the volume around a single particle from which all other particles are excluded. The attraction between molecules reduces the pressure by $N^2 a/V^2$, where $a$ accounts for the attractive potential of a particle due to the existence of all other particles. While for molecular gases, values of $a$ and $b$ can only be inferred from macroscopic measurements, for our colloidal system, we can determine the parameters $a$ and $b$ directly from the measured particle pair potential and particle size using $a = -2\pi \int_{r_1}^{\infty} U(r) r^2 dr$ and $b \approx 0.5 V_{excl}$ [14], where the excluded volume $V_{excl} = (4/3)\pi(2r_0)^3$. These relations can be strictly derived for particles interacting with a hard-core repulsion, followed by an attractive potential part for $r > r_1$, by summing the interactions over all particle pairs. Furthermore, while for molecular gases, the value of $a$ is temperature-independent, and the only change with temperature occurs in the thermal energy, $k_B T$, for our colloidal system, temperature changes directly the value of $a$ via the critical Casimir interactions. We determine values of $a$ by numerical integration of the measured pair potentials (table 1), and $b$ from the measured particle radius, and plot the resultant $P$ as a function of $V/Nb$ in Fig. 3. The values of $a$ result in four isotherms. Red and green curves show isotherms characteristic of a gas, while blue and pink curves show non-monotonic functions indicating gas-liquid condensation. To pinpoint the equilibrium gas-liquid coexistence, we construct horizontal tie lines that include equal areas with the isotherms [16, 17]; these tie lines intersect the isotherms at $v_{liq}$ and $v_{gas}$, the specific volumes of equilibrium liquid and gas phases that bound the coexistence regime. The model predicts that systems with average specific volume $\bar{v}$ within the coexistence regime separate into phases with relative fraction $f_{liq} = (v_{gas} - \bar{v})/(v_{gas} - v_{liq})$ and $f_{gas} = (\bar{v} - v_{liq})/(v_{gas} - v_{liq})$. To test this prediction, we determine the average volume per particle using $\bar{v} = b/4\phi$, and indicate the resulting value $\bar{v} = 12.5b$ with a vertical dashed line. This value falls within the coexistence regime of only the pink isotherm, indicating that gas-liquid coexistence should occur only for $\Delta T = 0.30^0 C$, in agreement with our experimental observation (Fig. 1). Further confirmation of the model comes from comparison of the measured and predicted values of $v_{liq}$ and $v_{gas}$. We determine the specific volumes of gas and liquid particles from the measured volume fractions $\phi_{liq}$=0.25 and $\phi_{gas}$=0.005 and obtain $v_{liq}/b$=1 and $v_{gas}/b$=50, in good agreement with the predicted values $v_{liq}/b$=1.25±0.31 and $v_{gas}/b$=43±25 determined from Fig. 3. Here, the error margin of the Van der Waals isotherms has been estimated from the error margin of the values of *a* and *b* due to the inaccuracy of the measured particle size, and the measured pair potential. Furthermore, the relative amount of liquid and gas is $f_{liq}$=0.69 and $f_{gas}$=0.31, again in good agreement with the values $f_{liq}$=0.72±0.1 and $f_{gas}$=0.28±0.1 determined from Fig. 3. These results indicate that the Van der Waals model provides a quantitatively accurate description of the colloidal gas-liquid equilibrium. We note that explicit measurement of the surface and bulk free energies [18] indicate that the critical radius of nucleation is two to three times smaller than the radius of the droplets analyzed here; this suggests that indeed the bulk free energy term starts to dominate the surface term, and the Van der Waals theory for bulk gas-liquid coexistence might apply.

The strength of the Van der Waals model is its universality, as reflected in the principle of corresponding states: A wide range of molecular gases shows the same behavior when the parameters pressure, temperature and density are scaled by their values at the critical point. We can test for this principle by comparing the normalized gas and liquid densities in our colloidal system with the well-known universal behavior of molecular gases. To do so, we determine the critical temperature $T_c=(8a/27bR)$ and critical density $\rho_c=N/3b$ using the Van der Waals parameters *a* and *b*, and the universal gas constant *R*. We plot the rescaled temperature $T/T_c$ as a function of the rescaled gas-liquid density difference in Fig. 4, inset. Here, we have taken $T$~325K, the temperature of the experiment. Circles denote the gas-liquid equilibrium described above, and squares indicate gas-liquid equilibria obtained for a binary solvent with composition slightly closer to the critical composition. The solid line indicates the universal behavior of a wide range of molecular gases [16]. This line provides a very good fit to the colloidal data without any adjustable parameters, indicating that this universal description applies to the colloidal system as well. We can further test the correspondence between our colloidal and molecular systems by using the law of rectilinear diameters for the average density [19,20] to determine the gas and liquid phase boundaries. We compare the rescaled temperature as a function of rescaled density of the colloid (symbols) with the universal behavior of molecular gases (solid lines) in the main panel of Fig. 4. Excellent agreement is observed, suggesting close correspondence between the colloidal system and molecular gases; this indicates that indeed the principle of corresponding states applies to the colloidal systems as well. We note that, however, more data close to the gas-liquid critical point is needed to strictly verify the corresponding critical scaling in the colloidal system.

Molecular gas-liquid condensation typically entails measurement of the equilibrium vapor pressure, $P_{eq}(T)$. For our colloidal system, this pressure is small and difficult to measure. However, we can estimate $P_{eq}$ from the intersection of the horizontal tie line with the y-axis. From Fig. 3, we find $P_{eq}=0.025(k_BT/b)$ yielding an equilibrium vapor pressure of $2 \cdot 10^{-4}$ Pa, $10^9$ times smaller than typical vapor pressures of molecular gases at room temperature, which are of the order of $\sim 10^5$ Pa, the atmospheric pressure. This difference reflects the $10^9$ times lower colloidal particle density associated with the $\sim 10^3$ times larger colloidal particle diameter. Therefore, the gas-liquid transitions in the colloidal and molecular systems are comparable.

The active and reversible control of colloidal gas-liquid and liquid-solid transitions lays the groundwork for novel assembly techniques making use of critical Casimir forces by a unique procedure of precise temperature control. Because of the reversibility of the interactions, temperature gradient and zone melting techniques can be applied to grow perfect equilibrium structures, in analogy to atomic crystal growth. Furthermore, the correspondence of the colloidal and the molecular gas-liquid transition opens new opportunities to further investigate equilibrium and non-equilibrium phenomena using active potential control. For example, quenches to temperatures even closer to the critical point induce fractal aggregates [21] and gelation of the particles. The temperature dependence in this case offers convenient control to tune the morphology of the assembled structures. Finally, the close agreement between the molecular Van der Waals theory and the colloidal phase separation that we observe on the scale of just a few particle diameters is surprising, and suggests that such mean field theories may also be applied to describe the condensation of nano particles in the assembly of nanostructures. We foresee that more complex structures can be obtained by patterning the particle surfaces to create anisotropic critical Casimir interactions. Such surface modification might allow mimicking "atomic orbitals" to assemble colloidal particles into structures similarly complex as those of molecules.

**Materials and Methods**

We use a quasi two-component solvent consisting of 3-Methyl Pyridine (3MP), water and heavy water [6, 22] with mass fractions of 0.25, 0.375, and 0.375, respectively. This composition is close to, but slightly lower than the critical composition with 3MP mass fraction 0.31 [23], resulting in a slightly extended temperature interval of aggregation. This solvent mixture separates into 3MP-rich and water-rich components upon heating to $T_c = 52.2°C$, the solvent phase separation temperature. This solvent was used for all graphs presented in this manuscript. In addition, in Fig. 4, we also include data from a binary solvent with 3MP mass fraction 0.28, in order to demonstrate a more diverse collection of data. To follow the aggregation due to critical Casimir forces deep in the bulk of the suspension, we use particles the refractive index and density of which match closely that of the solvent. Similar observations obtained with other particle materials show that the results presented here are general, and do not depend on the specific particle material used. We use poly-n-isopropyl acrylamide (PNIPAM) particles [24,25] that swell in the binary solvent. The swelling adjusts the particle refractive index and density to match closely that of the solvent in the swollen state, preventing particle sedimentation, and allowing direct observation of individual particles deep in the bulk. The particle size changes only weakly with temperature, and is constant in the temperature interval of interest here. At $T \sim T_c$, the particle radius is $r_0 \sim 250$ nm, and the particle volume fraction is ~2%. The particles are labeled with a fluorescent dye that makes them visible under fluorescent imaging. We tune critical Casimir interactions continuously by varying the solvent correlation length with temperature in a range of $\Delta T = 0.5$ to $0.2°C$ below $T_c$. We then follow colloidal phase transitions directly in real space and time by imaging individual particles in a 65μm by 65μm by 25μm volume with confocal microscopy, and determining their positions with an accuracy of ~0.03μm in the horizontal, and 0.05μm in the vertical direction [26].

**Acknowledgement**

We thank P. Bolhuis and D. Bonn for helpful discussions. This work was supported by the Innovational Research Incentives Scheme ("VIDI" grant) of the Netherlands Organization for Scientific Research (NWO) (P.S.).

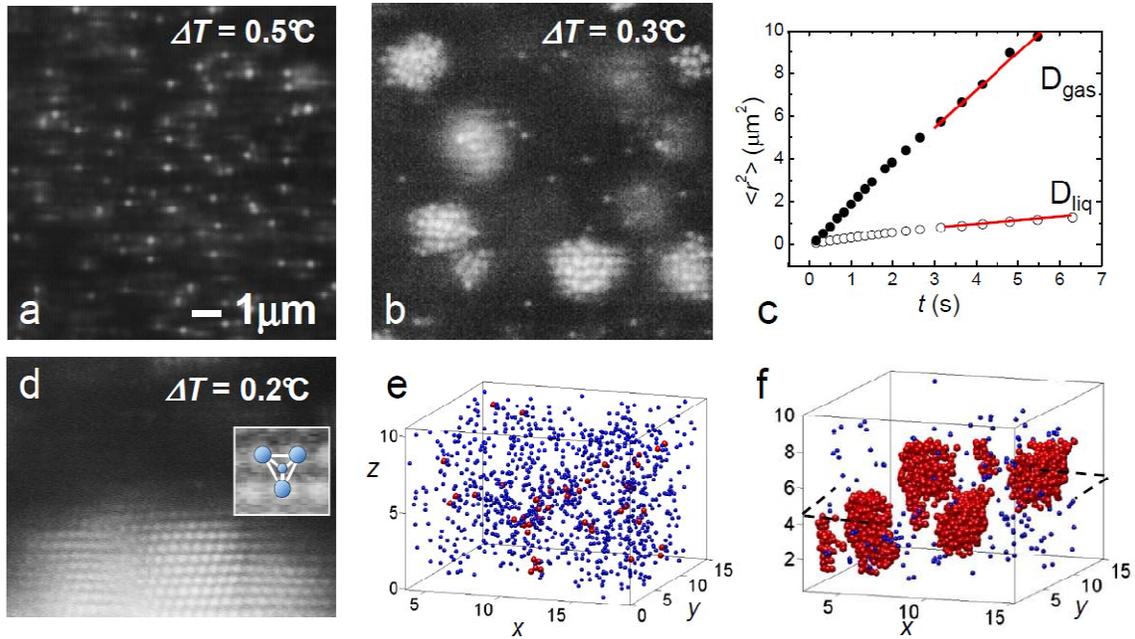

**Figure 1: Colloidal phase transitions induced by critical Casimir forces**

(a, b, d) Confocal microscope images of colloidal gas-liquid-solid transitions: colloidal gas (a), colloidal liquid aggregates (b), and colloidal crystal (d). Inset in (d) illustrates the relation between the imaged particle configuration and the structural motif of the face-centered cubic lattice. (c) Mean-square displacement of particles in the liquid (circles) and gas (dots). Linear fits to the data yields diffusion coefficients that differ by a factor of 10. (e,f) Three-dimensional reconstructions of gas and liquid correspond to the confocal images (a) and (b). Red and blue spheres indicate particles with a large and small number of nearest neighbors, respectively. Red particles in (e) demarcate density fluctuations of the colloidal gas, while red particles in (f) indicate the nucleated colloidal liquid phase coexisting with the colloidal gas.

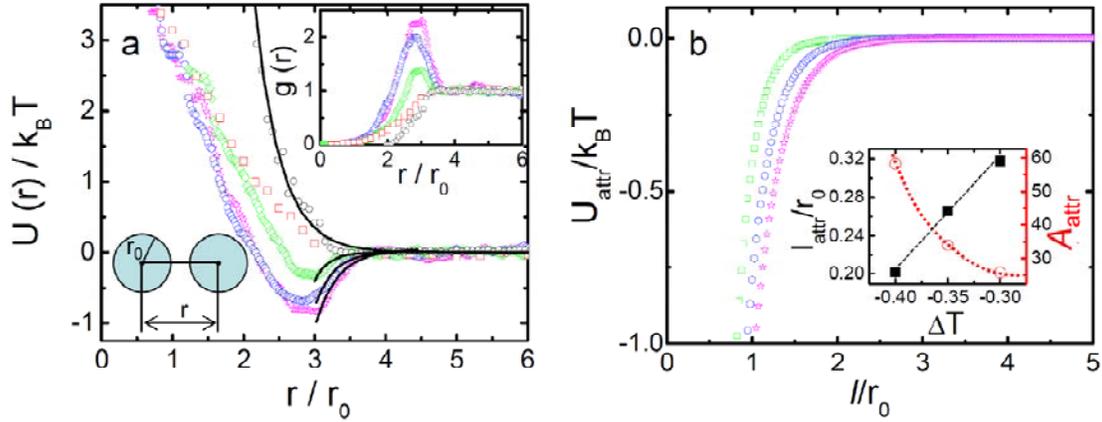

**Figure 2: Measurement of the particle pair potential**

a, Temperature-dependent particle pair potential determined from the pair correlation function shown in the inset. The symbols indicate room temperature (black circles), $\Delta T = 0.5\ ^0C$ (red squares), $0.4\ ^0C$ (green rhombi), $0.35\ ^0C$ (blue hexagons) and $0.30\ ^0C$ (magenta stars). The solid black line indicates the best fit with a screened electrostatic potential. (b) Attractive critical Casimir potentials extracted from the potential curves in (a). Inset: range $l_{attr}$ (black filled square) and amplitude $A_{attr}$ (red open circle) of the critical Casimir potential as a function of $\Delta T$. Dashed lines are guides to the eye.

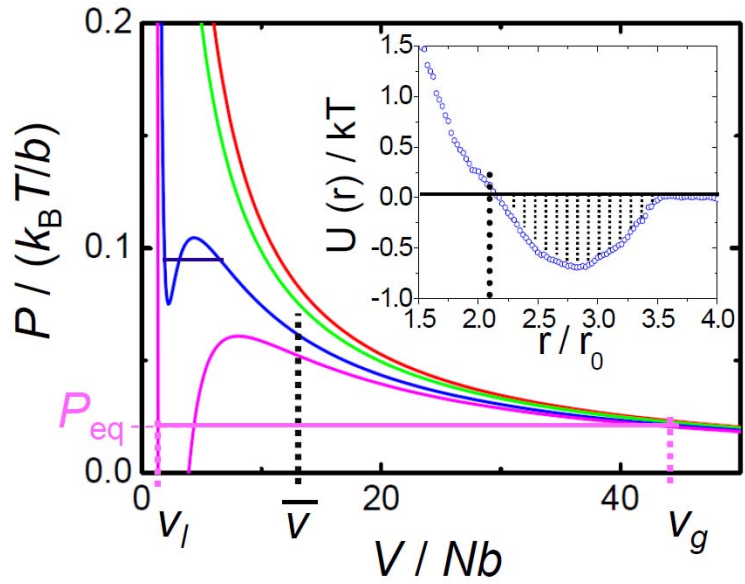

**Figure 3: Van der Waals isotherms of the colloidal gas-liquid transition.**

Van der Waals isotherms determined from the particle pair potential at $\Delta T = 0.5\,°C$ (red), $0.4\,°C$ (green), $0.35\,°C$ (blue), and $0.3\,°C$ (pink). A transition from a gas (green and red isotherms) to gas-liquid coexistence (blue and pink isotherms) is predicted. Horizontal tie lines are constructed according to the Maxwell rule of equal chemical potential, including equal areas with isotherms. Intersections with the isotherms delineate the gas-liquid coexistence regimes. Vertical dashed line indicates the average volume per particle, $\bar{v}$. Inset: Enlarged section of the particle pair potential at $\Delta T = 0.35\,^0C$ illustrates the determination of the Van der Waals parameter $a$ by integration. The lower integration boundary, $r_1$, is indicated by a dotted line.

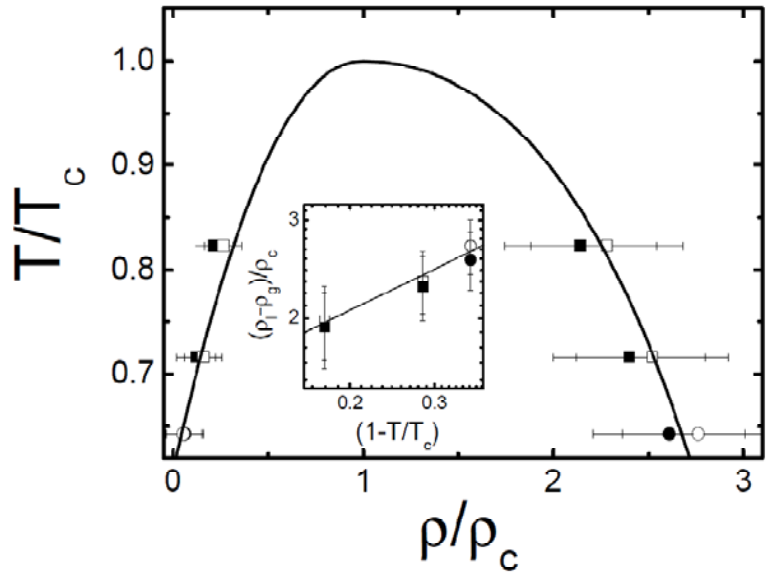

**Fig. 4 Colloidal gas-liquid phase diagram and comparison with molecular gases**

Normalized temperature as a function of normalized gas and liquid densities for the colloidal system (symbols) and for molecular gases (solid line). Open symbols indicate gas and liquid densities measured by microscopy, and closed symbols indicate the values obtained from the Van der Waals isotherms as shown in Fig. 3. The graph compiles data for two different binary solvents with 25% 3MP (circles) and 28% 3MP (squares); the latter extends the range of the colloidal interaction. The solid line indicates the universal behavior of a wide range of molecular gases according to the equation $[(\rho_{liq}-\rho_{gas})/\rho_c] = (7/2)[1-(T/T_c)]^{(1/3)}$ [16], and the law of rectilinear diameters [19,20]. It provides an excellent fit to the colloidal data. Inset: Normalized density difference of liquid and gas as a function of normalized temperature difference to the critical temperature. The solid line indicates the universal equation for molecular gases.

| T (°C) | $\Delta T$ (°C) | a ($bk_B T$) |
|---|---|---|
| 51.70 | 0.5 | 0 |
| 51.80 | 0.4 | 1.28 |
| 51.85 | 0.35 | 3.67 |
| 51.90 | 0.3 | 5.25 |

**Table 1 Temperature-dependent Van der Waals coefficient**

Values of the Van der Waals coefficient, *a*, for several temperatures. The values are determined by integration of the measured particle pair potential (see Fig. 3, inset).